\pdfoutput=1

\documentclass[aps,prd,twocolumn,superscriptaddress,showpacs]{revtex4}
\usepackage{graphics}
\usepackage{epstopdf}
\usepackage{graphicx}
\usepackage{dcolumn}
\usepackage{bm}
\usepackage{amsmath}
\usepackage{color}

\newcommand{\be}{\begin{equation}}
\newcommand{\ee}{\end{equation}}
\newcommand{\ba}{\begin{eqnarray}}
\newcommand{\ea}{\end{eqnarray}}
\newcommand{\bfig}{\begin{figure}[t]\begin{centering}}
\newcommand{\efig}{\end{centering}\end{figure}}

\begin{document}

\title{Preface: High Energy Astrophysics}

\author{Bing Zhang}
\email{zhang@physics.unlv.edu}
\affiliation{Kavli Institute for Astronomy and Astrophysics
and Department of Astronomy, Peking University, China;
Department of Physics and Astronomy, University
of Nevada, Las Vegas, USA}

\author{Peter M\'esz\'aros}
\email{nnp@astro.psu.edu}
\affiliation{Department of Astronomy and Astrophysics, Department of 
Physics, and Center for Particle and Gravitational Astrophysics,
Pennsylvania State University, USA}

\date{\today}

\begin{abstract}
High energy astrophysics is one of the most active branches in the
contemporary astrophysics. It studies astrophysical objects that
emit X-ray and $\gamma$-ray photons, such as accreting super-massive
and stellar-size black holes, and various species of neutron 
stars. With the operations of many space-borne and
ground-based observational facilities, high energy astrophysics has
enjoyed rapid development in the past decades. It is foreseen that the
field will continue to advance rapidly in the coming decade, with possible
ground-breaking discoveries of astrophysical sources in the high-energy 
neutrino and gravitational wave channels. This Special Issue of Frontiers 
of Physics is dedicated to a systematic survey of the field of high energy 
astrophysics as it stands in 2013. 
\end{abstract}                            
 
\maketitle


High energy astrophysics is the branch of astrophysics that studies 
astrophysical objects that emit high energy photons (X-rays and $\gamma$-rays),
and more generally, emit non-thermal
photons outside the traditional optical wavelengths.
By studying non-thermal emission in the universe, one can unveil
the part of universe that is not in a steady state, usually originating in
violent environments near compact objects, such as neutron stars and
black holes of different scales. These objects, besides emitting the
broad-band non-thermal electromagnetic radiation, are also believed to
be emitters of other signals outside the electromagnetic channel. 
These multi-messenger signals include cosmic rays, neutrinos, and
gravitational waves. The field of high energy astrophysics, along 
with cosmology and planetary sciences, is one of the three most active 
branches in the contemporary astrophysics. 

During the past decades, the field of high energy astrophysics has enjoyed
a rapid development, thanks to an impressive list of space-borne and
ground-based observational facilities. The currently operating space
missions dedicated to high energy astrophysics include the Chandra X-ray 
Observatory, the X-ray Multi-Mirror Mission (XMM)-Newton, Suzaku, and 
the Nuclear Spectroscopic Telescope Array (NuSTAR) in the X-ray
band; Swift and Integral in the hard X-ray/soft $\gamma$-ray band;
the Fermi Gamma-Ray Space Telescope and Astro-rivelatore Gamma a 
Immagini LEggero (AGILE) in the hard $\gamma$-ray band;
as well as Cosmic Ray Energetics and Mass (CREAM) and Alpha Magnetic 
Spectrometer Experiment (AMS-02) cosmic ray detectors. On the ground,
high energy facilities include the TeV telescopes High Energy 
Stereoscopic System (H.E.S.S.) and Very Energetic Radiation Imaging
Telescope Array System (VERITAS); cosmic ray detectors Pierre Auger
Observatory, {Telescope Array} and Astrophysical Radiation with Ground-based 
Observatory at YangBaJing (ARGO-YBJ); high energy neutrino detectors the IceCube
Neutrino Telescope, Astronomy with a Neutrino Telescope and Abyss 
environment RESearch (ANTARES) and Cubic Kilometre Neutrino Telescope
(KM3NeT); as well as gravitational wave detectors Laser
Interferometer Gravitational Wave Observatory (LIGO) and 
Virgo interferometer.

These observational facilities have greatly advanced our understanding
of various high-energy phenomena, including stellar-scale black holes
in binary systems, supermassive black holes in Active Galactic 
Nuclei (AGNs), gamma-ray bursts (GRBs), a rich list of objects in a
neutron star zoo, as well as supernovae and their remnants. The next 
10 years will see a continued growth
of the field, as technology has allowed several most challenging detectors 
to meet the sensitivity to potentially detect signals hitherto only 
speculated about. These include high-energy
(above 10 TeV) neutrinos and gravitational waves. It is foreseen that
the field of high energy astrophysics will fully embrace its
``multi-messenger'' nature as commonly envisioned. 

This Special Issue of Frontiers of Physics is intended
to provide a comprehensive overview of this exciting field. Reviews of various 
individual topics are available from many sources. However, there is no 
single source where interested readers can find an 
overview of all the directions
in this field. This issue is aimed at filling this gap. The original plan
was to publish 12 articles in 3 parts. It turned out that one review
article (supernovae and their remnants) could not be finished in time,
so the current issue only includes 11 articles.

The first part of the Special Issue is designed to survey the field 
according to high energy objects or phenomena. It includes active 
galactic nuclei (AGNs, H. Krawczynski \& E. Treister) \cite{Krawczynski13}
that host super-massive black holes, stellar-size black hole X-ray 
binaries (BHXBs, S.-N. Zhang) \cite{Zhang13},  gamma-ray bursts
(GRBs, N. Gehrels \& S. Razzaque) \cite{Gehrels13}, and various 
species in the neutron star (NS) zoo (A. K. Harding). The 5th 
planned article on supernovae and their remnants did not materialize.

In \cite{Krawczynski13}, Krawzynski and Treister review the astrophysics
of AGNs, the nuclei of galaxies that host super-massive black
holes in the mass range $10^6-10^{10} M_\odot$, which accrete material
from the environment. In some cases, these accreting BHs launch a pair
of powerful jets. Depending on the viewing angle to these systems and to 
a lesser degree on the intrinsic properties as well, AGNs display themselves 
in a rich trove of phenomenology. These are reviewed along with the underlying
physics of jets and central engine, as well as evolution of AGNs within the
cosmological context.

In \cite{Zhang13}, S.-N. Zhang surveyed our current understanding of 
black hole X-ray binaries (BHXBs). These
systems include a stellar-mass black hole, typically with a mass greater
than $(3-20) M_\odot$, and a companion star from which matter is accreted 
towards the black hole (BH) and form an accretion disk 
which is hot enough to emit X-rays. Jets
are often associated with these systems, so these objects are also termed
as ``micro-quasars'' by analogy with AGNs. The phenomenology and underlying
physics are reviewed, with the focus on several important subjects, such as
BH spin measurements, hot accretion flows and corona physics, as well as 
transition between two accretion states. 

In the extreme of stellar-scale high-energy events, GRBs distinguish themselves
as the most luminous explosions in the universe, which signal the birth of
stellar-mass black holes or rapidly spinning highly-magnetized neutron stars (NSs). 
This subject is reviewed by Gehrels and Razzaque \cite{Gehrels13}, with the focus
on the recent observational breakthroughs led by the NASA missions Swift
and Fermi. These observations have greatly `advanced our understanding of the 
physics of the GRB phenomenology, including the origin of the prompt gamma-ray emission
and afterglow, as well as the physics of the central engine and the progenitor
star of GRBs.

Besides BHs, another type of objects that high energy astrophysics studies
extensively are neutron stars. These objects are believed to be composed of
matter whose density is close to the nuclear density but whose interior
composition remains unknown. There is a diverse population of NSs. In 
\cite{Harding13}, Harding presents a comprehensive review of the observational 
and physical properties of the different species in the NS zoo, as well as 
the physical origins of the diversity. The objects surveyed include
rotation-powered pulsars, magnetars, compact central objects, isolated
neutron stars, and accreting neutron stars.

The second part of the Special Issue surveys the field of high energy
astrophysics by means of the detection methods. There are many detectors
that cover different observational channels (as summarized in the 2nd 
paragraph). It is impossible to review here all of them. With the focus on
the highest energy and non-electromagnetic channels, we selected the following
5 topics: $\sim$ GeV $\gamma$-ray astronomy examplified by Fermi (S. Ritz) 
\cite{Ritz13}{\bf ;}  $\sim$ TeV $\gamma$-ray astronomy (F. M. Rieger, E.
de Ona-Wilhelmi, and F. A. Aharonian) \cite{Rieger13}{\bf ;} high energy cosmic ray
astronomy (T. K. Gaisser, T. Stanev, and S. Tilav) \cite{Gaisser13}{\bf ;} high 
energy neutrino astronomy (F. Halzen) \cite{Halzen13}{\bf ;} and 
gravitational wave astronomy (G. Gonzalez, A. Vicer\'e, and L. Wen) 
\cite{Gonzalez13}.

The Fermi satellite has been extremely successful in unveiling the high
energy universe from 30 MeV to $>$ 300 GeV. In \cite{Ritz13}, Ritz summarizes
the achievements of Fermi in various disciplines of astrophysics, including
pulsars and nebulae, AGNs, GRBs, the ``Fermi bubbles, and the $\gamma$-ray 
background. The endeavors of using Fermi data to shed light on fundamental
physics problems, such as indirect search for dark matter signals as well
as constraints on Lorentz Invariance Violation, are also highlighted.

The TeV astronomy domain was recently rapidly advanced thanks to the 
successful completion of several ground-based detectors, such as H.E.S.S. and 
VERITAS. The exciting discoveries  at TeV energies are reviewed by
Rieger et al. \cite{Rieger13} along with their astrophysical implications. 

Compared with other branches, cosmic ray astronomy may be the earliest
branch in high energy astrophysics. Yet, the energy spectra of all particles
and different species are still not measured precisely, and their physical
origins remain elusive. In \cite{Gaisser13}, Gaisser et al. critically 
review this field, paying special attention to the energy spectrum in
the PeV range and above. They present a baseline spectrum from 
$10^{14}$ to $10^{20}$ eV obtained by combining the measurements published in
various sources, and they discuss the astrophysical implications of the
results.

High energy neutrinos are believed to be produced through hadronic
interaction processes from astronomical sources that accelerate high
energy cosmic rays. Detecting them is extremely challenging. In
\cite{Halzen13}, Halzen reviews the journey of constructing the first
kilometer-scale neutrino detector IceCube in Antarctica, discussing the
astrophysical motivations to construct such a detector, as well as
the prospects to detect high energy neutrinos from astrophysical 
sources in the near future.

In another direction, Gonzalez et al. \cite{Gonzalez13} show that gravitational
wave astronomy is entering a new era in which the very first direct detection
of gravitational waves becomes possible. They review various astrophysical
sources that are likely gravitational wave emitters, and show how the current
technology is getting close to reaching the sensitivity to detect these
faint signals. 

The Special Issue ends with the third part that includes two reviews on
two important subjects of the contemporary astrophysics, namely, dark 
matter (X.-J. Bi, P.-F. Chen and Q. Yuan) \cite{Bi13} and dark energy 
(M. Li, X.-D. Li, S. Wang, Y. Wang) \cite{Li13}. 
These two subjects are not limited to high energy astrophysics, having
a large impact on astrophysics as whole (especially cosmology), 
as well as being major problems of fundamental physics. Nonetheless, 
they are closely connected to high energy astrophysics. For example,
observations of electron/positron cosmic rays, $\gamma$-ray lines,
and neutrinos may lead to indirect detection of dark matter. The
existence of dark energy was first revealed by observing the space distribution
of explosions of high-energy events Type Ia supernovae, and the nature 
of dark energy may be revealed by
combining measurements of different standard candles, including
those in high energy phenomena such as GRBs.

In \cite{Bi13}, Bi et al. review the status and progress in dark
matter (DM) searches, including direct searches of recoils and ionization
from DM-kicks to standard model (SM) particles in underground experiments, 
collider searches of missing energy carried by DM particles, as well as
indirect searches of $\gamma$-rays and leptonic cosmic rays due
to DM-DM interactions in astrophysical environments. Even though no
robust detection is available yet, they show several 
interesting observations that may suggest evidence of DM,
and they discuss the promises and issues of these observations.

Finally, Li et al. \cite{Li13} review various astrophysical lines of
evidence for the existence of dark energy (DE), and an impressive
list of theoretical particle physics models that interpret the DE
phenomenon. They also discuss how the theoretical models are 
constrained by the available and future astrophysical data.

We are lucky to live in a special epoch in the human history. 
Observations have allowed us to measure the parameters of our universe 
with increasing precision, and to infer the composition of universe
we are living in. Ground-breaking discoveries over the decades have extended
our horizon to understand the universe, from the brightest
(e.g. GRBs, AGN blazars) to the darkest (e.g. dark matter and dark
energy). More excitingly, we are on the technological brink of
observing the universe using brand new observational channels,
such as high energy neutrinos and gravitational waves. For the first
time in history, the next decade will probably witness the first truly
multi-messenger observational campaign of a high energy phenomenon.
Theoretical modeling will be advanced to address these new discoveries,
with many open questions answered or partially addressed. Numerical
simulations will continue to ripen by combining general relativity and
resistive MHD at the macroscopic level and particle-in-cell simulations
at the microscopic level. More advanced computers and algorithms will
allow simulations to unprecedented levels of accuracy to address
fundamental questions such as magnetospheric structure, 
jet launching, energy dissipation, particle acceleration and 
radiation in high energy astronomical systems.
If history is any guide, it is foreseen that new observations
will raise more questions and challenges, which will give continuous
impetus for the further development of new technologies and theoretical 
ideas to better understand the exciting and mysterious universe
we are living in.

Finally, we'd like to thank all the invited reviewers of this
Special Issue, without whose dedication and great efforts this
volume would not be possible. We also thank Mr. Hong-Guang Dong
for restlessly providing editorial support to the authors.


\begin{thebibliography}{11}

\bibitem{Krawczynski13}
H. Krawczynski, E. Treister, Frontier of Phys. 2013, 8(6): 609, 
arXiv:1301.4179

\bibitem{Zhang13}
S.-N. Zhang, Frontier of Phys. 2013, 8(6): 630, arXiv:1302.5485

\bibitem{Gehrels13}
N. Gehrels, S. Razzaque, Frontier of Phys. 2013, 8(6): 661,
arXiv:1301.0840

\bibitem{Harding13}
A. K. Harding, Frontier of Phys. 2013, 8(6): 679, arXiv:1302.0869

\bibitem{Ritz13}
S. Ritz, Frontier of Phys. 2013, 8(6): 693

\bibitem{Rieger13}
Rieger, F. M., de Ona-Wilhelmi, E., Aharonian, F. A.
rontier of Phys. 2013, 8(6): 714, arXiv:1302.5603

\bibitem{Gaisser13}
Gaisser, T. K., Stanev, T., Tilav, S. 
Frontier of Phys. 2013, 8(6): 748, arXiv:1303.3565

\bibitem{Halzen13}
Halzen, F. Frontier of Phys. 2013, 8(6): 759 

\bibitem{Gonzalez13}
Gonzalez, G., Vicer\'e, A., Wen, L. Frontier of Phys. 2013, 8(6): 771

\bibitem{Bi13}
Bi, X.-J., Yin, P.-F., Yuan, Q. Frontier of Phys. 2013, 8(6): 794

\bibitem{Li13}
Li, M., Li, X.-D., Wang, S., Wang, Y.
Frontier of Phys. 2013, 8(6): 828, arXiv:1209.0922










\end{thebibliography}
\end{document}